\documentclass[journal,10pt,twocolumn,nofonttune]{IEEEtran}
\usepackage{xmpaper,lzqsymbol}
\usepackage{amssymb,amsmath,latexsym}
\usepackage{graphicx,subfigure}
\usepackage{color,cite,times}
\usepackage{epstopdf}
\usepackage{multirow}
\usepackage{array}

\usepackage{amsmath}
\usepackage{bm}
\usepackage[T1]{fontenc}

\usepackage{booktabs}
\usepackage{bbm} 
\usepackage{subfigure}

\begin{document}
\title{ Passive Beamforming and Information Transfer via Large Intelligent Surface}

\author{Wenjing Yan, Xiaoyan Kuai, and
	Xiaojun~Yuan,~\IEEEmembership{Senior~Member,~IEEE}\vspace{-0.5cm}
\thanks{W. Yan, Y. Kuai, and X. Yuan are with the Center for Intelligent Networking and Communications, the National Laboratory
of Science and Technology on Communications, the University of Electronic Science and Technology of China, Chengdu 611731,
China (e-mail: wjyan@std.uestc.edu.cn; xy$\_$kuai@uestc.edu.cn; xjyuan@uestc.edu.cn.).
}
}

\maketitle
\begin{abstract}
  Large intelligent surface (LIS) has emerged as a promising new solution to improve the energy and spectrum efficiency of wireless networks. A LIS, composed of a large number of low-cost and energy-efficient reconfigurable passive reflecting elements, enhances wireless communications by reflecting impinging electro-magnetic waves. In this paper, we propose a novel passive beamforming and information transfer (PBIT) technique, in which the LIS simultaneously enhances the primary communication and sends information to the receiver. We develop a passive beamforming method to improve the average receive signal-to-noise ratio (SNR). We also establish a two-step approach at the receiver to retrieve the information from both the transmitter and the LIS. Numerical results show that the proposed PBIT system, especially with the optimized passive beamforming, significantly outperforms the system without LIS enhancement. Furthermore, a tradeoff between the passive-beamforming gain and the information rate of the LIS has been demonstrated.
\end{abstract}

\section{Introduction}

Recent years have witnessed an explosive growth of wireless data demands along with the popularity of smart terminals and mobile devices \cite{boccardi2013five}. Although the utilization of advanced wireless technologies, such as millimetre wave (mmWave), massive multiple-input multiple-output (MIMO), ultra-dense deployments, etc., has greatly improved the spectral efficiency of wireless networks\cite{andrews2014will}, the resulting energy consumption and hardware cost problems have become a bottleneck restricting the practical implementation of these technologies\cite{zhang2017fundamental}. To reduce the energy consumption and improve the energy efficiency of wireless networks, large intelligent surface (LIS)\cite{nadeem2019large}, a.k.a. intelligent reflecting surface (IRS)\cite{wu2018intelligent} has been envisioned as a promising new hardware solution to enhance future wireless communication systems. A LIS is composed of a large number of low-cost and energy-efficient reconfigurable reflecting elements that can reflect impinging electromagnetic waves with a controllable phase shift via the help of a smart controller. It is worth noting that passive reflecting surfaces have various applications in radar and satellite communications, but has been rarely used in terrestrial wireless communications. The reason is that traditional reflecting surfaces only have fixed phase shifters and cannot adapt to the time-varying environment in terrestrial communications. However, with the recent developments in metasurfaces, reconfiguration of reflecting surfaces is now made possible via controlling the phase shifters in real time \cite{hum2014reconfigurable,foo2017liquid}.
 As such, passive beamforming, in which the phase shifts of the reflecting elements of a LIS are intelligently adjusted to achieve coherent superposition of the reflected signals at a desired receiver, has been studied in \cite{nadeem2019large, wu2018intelligent, huang2018large}  to substantially enhance the energy efficiency of wireless communications.

\begin{figure}
  \centering
  \includegraphics[width=3.5 in]{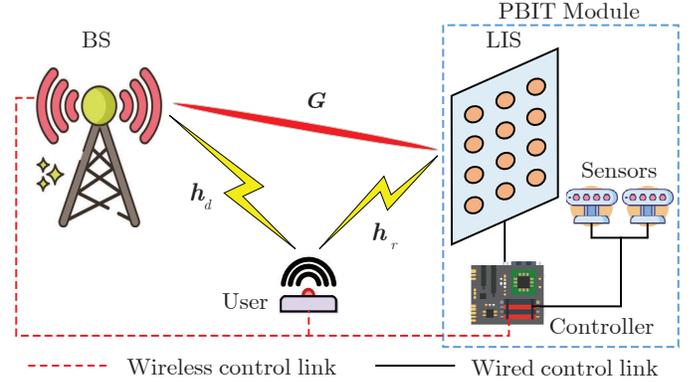}\\
  \caption{A PBIT-enhanced wireless system based on LIS.}\label{IM}
\end{figure}
In this paper, we propose a novel passive beamforming and information transfer (PBIT) enhanced wireless system, where a signal-antenna user communicates with a multi-antenna base station (BS) through the help of a LIS, as illustrated in Fig. 1. Compared to the work in \cite{nadeem2019large, wu2018intelligent, huang2018large}, a major difference of the PBIT system is that, besides performing passive beamforming to enhance the user-BS communication, the LIS is also required to transmit data to the receiver. There are a variety of potential sources for the LIS data, e.g., low-cost sensors implemented in a smart building for collecting environmental data such as temperature, humidity, tension, etc. The challenge then resides in the design of the LIS operations to simultaneously transmit data and enhance the user-BS communications via passive beamforming, as well as the design of the receiver operations to retrieve both the information from the transmitter and the LIS.

The main contributions of this paper are as follows. First, we propose to use spatial modulation\cite{mesleh2008spatial} for LIS data, i.e., the LIS information is carried by the on/off states of the LIS reflecting elements\footnote{The state ``off'' of a reflecting element means that there is only structure-mode reflection generated as if the element is a regular conductor. The structure-mode reflection can be absorbed into the direct link channel in channel modeling\cite{zhang2019constellation}. The state ``on'' means that there are
both structure-mode reflection and antenna-mode reflection, in which the load impedance mismatches the antenna impedance\cite{green1963general}.},
 while passive beamforming is achieved by adjusting the phase shifts of the activated reflecting elements. Second, in passive beamforming design, we formulate the problem of maximizing the average receive signal-to-noise ratio (SNR), and use the semidefinite relaxation (SDR) technique to obtain a suboptimal solution to the problem. Third, to retrieve both the information from the transmitter and the LIS, we develop an efficient two-step detection algorithm involving compressed sensing and matrix factorization techniques. Substantial performance gains have been demonstrated for the optimized PBIT scheme. The tradeoff between the passive-beamforming gain and the information rate of the LIS has also been demonstrated.

\section{System Model and Problem Description}

\subsection{System Model}
\label{sec.sys}

Consider a PBIT enhanced wireless communication system, as illustrated in Fig.~\ref{IM}. The system combines a single-input multiple-output (SIMO) wireless communication system with a PBIT module,
where a single-antenna user communicates with a base station (BS) equipped with $M$ antennas.
The PBIT module consists of a LIS equipped with $N$ passive reflecting elements, a controller to adaptively adjust the on/off state and the phase shift of each passive reflecting element, and a number of sensors to collect the environmental data. The sensors send their collected data to the controller through a wired link. Then, the controller adjusts the on/off state of each passive reflecting element according to the sensor data.
Meanwhile, the activated reflecting elements reflect the signals transmitted from the user to enhance the user-BS communication.
The phase of the reflected signals can be adjusted by the controller to optimize the system performance.

We ignore the signal power reflected by the LIS for two or more times due to severe path loss.
Denote by $\boldsymbol{h}_d \in \mathbb{C}^{M\times 1}$, $\boldsymbol{h}_r \in \mathbb{C}^{N\times 1}$, and $\boldsymbol{G} \in \mathbb{C}^{M\times N}$ the baseband equivalent channels of the user-BS link, the user-LIS link, and the LIS-BS link, respectively, where $\mathbb{C}^{a \times b}$ is the space of $a \times b$ complex-valued matrices.
We assume that all the channel links are quasi-static and flat-fading.
Let $\boldsymbol{\Theta} = \text{diag}\{ \boldsymbol{\theta}\}$ denote the diagonal phase-shift matrix for the LIS,
where $ \boldsymbol{\theta} = [\theta_1, \theta_2, \ldots, \theta_N]^{\rm T}\in \mathbb{C}^{N\times 1}$ and $|\theta_n| = 1$. Define $\beta \in [0,1]$ as the amplitude reflection coefficient.
Let $s_i$ be the state of $i$th passive reflecting element, with $s_i=1$ meaning that the state of $i$th element is ``on'' and $s_i=0$ otherwise. Denote by $\boldsymbol{s} = [s_1, s_2, \ldots, s_N]^{\rm T}$ the state of LIS that carries the information from the sensors. We assume that each $s_n$ independently takes the value of $1$ (``on'') with probability $\rho$ and the value of $0$ (``off'') with probability $1-\rho$, i.e.,
\begin{align} \label{pro.s}
  p(\boldsymbol{s}) = \prod_{n=1}^N p(s_n) = \prod_{n=1}^N(1-\rho)^{1-s_n} \rho^{s_n}.
\end{align}
Then, each $s_n$ carries $H(\rho)=-\rho \log \rho - (1-\rho) \log (1-\rho)$ bit of information.

We assume block transmission with each transmission block consisting  of $L$ time slots. The observed signal at the receiver in the $l$th time slot is
\begin{align} \label{channel.H1}
 \boldsymbol{y}_l = (\beta\boldsymbol{G} \boldsymbol{\Theta}\boldsymbol{S}\boldsymbol{h}_r  + \boldsymbol{h}_d )x_l +
 \boldsymbol{w}_l,
\end{align}
where $\boldsymbol{y}_l \in \mathbb{C}^{M \times 1}$ is the observed signal vector, $\boldsymbol{S} = \diag\{\boldsymbol{s}\} \in \mathbb{R}^{N \times N}$ is a diagonal matrix,
$x_l\in \mathbb{C}$ is the transmit signal in the $l$th time slot,
and $\boldsymbol{w}_l\in \mathbb{C}^{M \times 1}$ is an additive white Gaussian noise (AWGN) with the elements independently drawn from $\mathcal{CN}(0, \sigma_w^2)$. We assume that the diagonal matrix $\boldsymbol{S}$ remain fixed over each transmission block.
Then, the observed signal matrix of a transmission block, denoted by $\boldsymbol{Y}=[\boldsymbol{y}_1, \ldots, \boldsymbol{y}_L]$, can be expressed as \begin{align} \label{channel.H2}
 \boldsymbol{Y} = (\beta\boldsymbol{G} \boldsymbol{\Theta}\boldsymbol{S}\boldsymbol{h}_r  + \boldsymbol{h}_d )\boldsymbol{x}^{\rm T} +
 \boldsymbol{W},
\end{align}
where $\boldsymbol{x}=[x_1,\ldots,x_L]^{\rm T}$ and $\boldsymbol{W}=[\boldsymbol{w}_1, \ldots, \boldsymbol{w}_L]$.

Each entry of $\boldsymbol{x}$ is modulated by using a constellation $\mathcal{C}=\{c_1,c_2,\ldots,c_{\vert \mathcal{C}\vert}\}$, where $\vert \mathcal{C}\vert$ is the cardinality of $\mathcal{C}$. That is, $x_l$ is uniformly drawn from $\mathcal{C} $ for $\forall l$, where $x_l$ is the $l$th entry of $\boldsymbol{x}$. Denote by $P$ the power budget at the user, i.e., $\frac{1}{L}\boldsymbol{x}^{\rm H}\boldsymbol{x} \leqslant P$.

In this paper, we assume perfect channel state information (CSI), i.e., the channel state $\{ \beta,\boldsymbol{G},\boldsymbol{h}_r,\boldsymbol{h}_d\}$ is perfectly known by the BS. With the available CSI, the BS is able to determine the optimal phase shifts $\boldsymbol{\Theta}$ of the LIS in the sense of certain design criteria specified in the next subsection. Then, the BS sends the optimal
$\boldsymbol{\Theta}$ to the LIS through a control link. The acquisition of the CSI can be done, e.g., by employing the recently developed channel estimation techniques in \cite{he2019cascaded} and \cite{taha2019enabling}. Details are omitted here due to space limitations.

\subsection{Problem Description}

In this paper, we aim to retrieve both the information from the user and the sensors (i.e., $\boldsymbol{x}$ and  $\boldsymbol{s}$) at the receiver with the help of the LIS. More specifically, given the information rate of $\boldsymbol{x}$ and $\boldsymbol{s}$, we need to design the phase-shift matrix $\boldsymbol{\Theta}$ such that the receiver is able to reliably recover $\boldsymbol{x}$ and $\boldsymbol{s}$ with a minimum transmission power $P$.

From information theory, the sum capacity of the PBIT system in \eqref{channel.H2} is given by the mutual information $I(\boldsymbol{x},\boldsymbol{s};\boldsymbol{Y})$\cite{cover2012elements}. Then, our design problem can be decoupled into two subproblems: one is the passive beamforming design, i.e., to maximize $I(\boldsymbol{x},\boldsymbol{s};\boldsymbol{Y})$ over the phase shift matrix $\boldsymbol{\Theta}$; and the other is the transceiver design, i.e., to design the signaling of $(\boldsymbol{x},\boldsymbol{s})$ and the receiver at the BS to achieve the obtained maximum $I(\boldsymbol{x},\boldsymbol{s};\boldsymbol{Y})$.

We first consider the passive beamforming design. $I(\boldsymbol{x},\boldsymbol{s};\boldsymbol{Y})$ is difficult to evaluate since \eqref{channel.H2} is a complicated model. To avoid this difficulty, we propose a heuristic design metric as follows.
 Note that the required rate of $\boldsymbol{x}$ is typically much higher than that of $\boldsymbol{s}$ in a practical scenario. Then,
\begin{subequations} \label{Inf.1}
\begin{align}
&I(\boldsymbol{x},\boldsymbol{s}; \boldsymbol{Y})= I(\boldsymbol{x};\boldsymbol{Y}|\boldsymbol{s}) + I(\boldsymbol{s};\boldsymbol{Y}) \label{Inf.1.a}\\
&~~~~~~~~~~~~~ \approx I(\boldsymbol{x};\boldsymbol{Y}|\boldsymbol{s}) \label{Inf.1.b}\\
&~~~~~~~~~~~~~ = \mathbb{E}\log(1 + \text{SNR}(\boldsymbol{s})) \label{Inf.1.c}\\
&~~~~~~~~~~~~~ \leq \log(1 + \mathbb{E} [\text{SNR}(\boldsymbol{s})]) \label{Inf.1.d}\\
&~~~~~~~~~~~~~ = \log\left(1 + \frac{\mathbb{E} \| \boldsymbol{G} \boldsymbol{\Theta}\boldsymbol{S}\boldsymbol{h}_r  + \boldsymbol{h}_d \|_2^2 P}{\sigma_w^2} \right)\label{Inf.1.e},
\end{align}
\end{subequations}
where \eqref{Inf.1.a} follows from the chain rule of mutual information; in \eqref{Inf.1.c},
$\text{SNR}(\boldsymbol{s}) \triangleq \frac{\| \boldsymbol{G} \boldsymbol{\Theta}\boldsymbol{S}\boldsymbol{h}_r  + \boldsymbol{h}_d \|_2^2 P}{\sigma_w^2}$ with $\|\cdot\|_2$ being the $\ell_2$-norm,  and the expectation $\mathbb{E}$ is taken over $\boldsymbol{s}$;  and \eqref{Inf.1.d} follows from the Jensen's inequality and the concavity of the logarithm function. Based on the above, we henceforth aim to maximize $\mathbb{E} \| \boldsymbol{G} \boldsymbol{\Theta}\boldsymbol{S}\boldsymbol{h}_r  + \boldsymbol{h}_d \|_2^2 $ in the passive beamforming design.

As for the transceiver design, we will focus on the design of the receiver at the BS to reliably recover both $\boldsymbol{s}$ and $\boldsymbol{x}$ from the received signal $\boldsymbol{Y}$ (for a given $\boldsymbol{\Theta}$). From information theory, besides the receiver design, we also need to design signal shaping and channel coding at the receiver, so as to approach the channel capacity.  These problems are, however, out of the scope of this paper.

\section{Beamforming Design}

From \eqref{Inf.1} and the discussions therein, the optimization of the phase shift $\boldsymbol{\theta}$ at the LIS can be formulated as
\begin{subequations} \label{optimal.2}
\begin{align}
& \max_{\boldsymbol{\theta}} \quad \mathbb{E} \|\boldsymbol{G} \boldsymbol{\Theta}\boldsymbol{S}\boldsymbol{h}_r  + \boldsymbol{h}_d \|_2^2 \label{optimal.2.a}\\
& ~~\textrm{s.t.} \quad~  |\theta_n| = 1, \text{for}~n = 1,\ldots, N. \label{optimal.2.b}
\end{align}
\end{subequations}
Let $\boldsymbol{D}_h = \diag\{\boldsymbol{h}_r\}$. Then,
\begin{align} \label{optimal.3}
&\mathbb{E} \|\boldsymbol{G} \boldsymbol{\Theta}\boldsymbol{S}\boldsymbol{h}_r  + \boldsymbol{h}_d \|_2^2  \notag\\
&=  \mathbb{E} \left[\boldsymbol{s}^{\rm H} \boldsymbol{D}_h^{\rm H} \boldsymbol{\Theta}^{\rm H}
\boldsymbol{G}^{\rm H} \boldsymbol{G} \boldsymbol{\Theta}\boldsymbol{D}_h \boldsymbol{s} +
2 \textrm{Re}(\boldsymbol{s}^{\rm H} \boldsymbol{D}_h^{\rm H} \boldsymbol{\Theta}^{\rm H}
\boldsymbol{G}^{\rm H} \boldsymbol{h}_d)  \right]  \notag\\
&= \textrm{tr}\left(\boldsymbol{G} \boldsymbol{\Theta}\boldsymbol{D}_h \mathbb{E}\left[\boldsymbol{s} \boldsymbol{s}^{\rm H}\right] \boldsymbol{D}_h^{\rm H} \boldsymbol{\Theta}^{\rm H}
\boldsymbol{G}^{\rm H}\right) \notag\\
&~~~+ 2 \textrm{Re}\left(\mathbb{E}\left[\boldsymbol{s}^{\rm H}\right] \boldsymbol{D}_h^{\rm H} \boldsymbol{\Theta}^{\rm H}
\boldsymbol{G}^{\rm H} \boldsymbol{h}_d \right),
\end{align}
where $\textrm{Re}(a)$ denotes the real part of the complex number $a$.
Based on the probability distribution of $\boldsymbol{s}$ in \eqref{pro.s}, we obtain
\begin{align} \label{optimal.4}
&\mathbb{E}\left[\boldsymbol{s} \boldsymbol{s}^{\rm H}\right] = \rho^2 \textbf{1}\cdot \textbf{1}^{\rm T} + \rho(1-\rho)\mathbf{I} \quad \text{and}
\quad \mathbb{E}\left[\boldsymbol{s}^{\rm H}\right] = \rho \textbf{1},
\end{align}
where $\textbf{1}$ is an $N$-dimension all-one vector, and $\mathbf{I}$ is the identity matrix with an appropriate size.
From the discussion below \eqref{pro.s}, $\rho$ is uniquely determined by the information rate of $\boldsymbol{s}$. For a given target rate $r$ of $\boldsymbol{s}$, we have $\rho = H^{-1}(r)$.
Plugging \eqref{optimal.4} into \eqref{optimal.3}, we obtain
\begin{align}  \label{optimal.6}
&\mathbb{E} \|\boldsymbol{G} \boldsymbol{\Theta}\boldsymbol{D}_h \boldsymbol{s}  + \boldsymbol{h}_d \|_2^2  \notag\\
& = \rho^2 \textbf{1}^{\rm H} \boldsymbol{D}_h^{\rm H} \boldsymbol{\Theta}^{\rm H}
\boldsymbol{G}^{\rm H} \boldsymbol{G} \boldsymbol{\Theta}\boldsymbol{D}_h \textbf{1} +2\rho \textrm{Re}(\textbf{1}^{\rm H} \boldsymbol{D}_h^{\rm H} \boldsymbol{\Theta}^{\rm H} \boldsymbol{G}^{\rm H} \boldsymbol{h}_d) \notag\\
&~~~~~~~~~ + \rho(1-\rho) \textrm{tr} \left(\boldsymbol{D}_h^{\rm H} \boldsymbol{\Theta}^{\rm H} \boldsymbol{G}^{\rm H} \boldsymbol{G} \boldsymbol{\Theta}\boldsymbol{D}_h \right)  \notag\\
& = \rho^2 \boldsymbol{\theta}^{\rm H} \boldsymbol{D}_h^{\rm H}
\boldsymbol{G}^{\rm H} \boldsymbol{G} \boldsymbol{D}_h \boldsymbol{\theta} + 2\rho \textrm{Re}(\boldsymbol{\theta}^{\rm H} \boldsymbol{D}_h^{\rm H}
\boldsymbol{G}^{\rm H} \boldsymbol{h}_d) \notag\\
&~~~~~~~~~~ + \rho(1-\rho) \boldsymbol{\theta}^{\rm H} \diag\{\boldsymbol{v}\}\boldsymbol{\theta}
\end{align}
where $\boldsymbol{v}$ is the diagonal of $ \boldsymbol{D}_h^{\rm H}\boldsymbol{G}^{\rm H} \boldsymbol{G} \boldsymbol{D}_h $.
With \eqref{optimal.6}, we see that the problem in \eqref{optimal.2} is a non-convex quadratically constrained quadratic program (QCQP).
Following \cite{wu2018intelligent} and \cite{zhang2017multi}, we approximate problem \eqref{optimal.2} as a semidefinite program (SDP). The details are presented below.

We first reformulate the optimization problem as a homogeneous QCQP by introducing an auxiliary variable $t$, yielding
\begin{subequations} \label{optimal.7}
\begin{align}
& \max_{\bar{\boldsymbol{\theta}}} \quad \bar{\boldsymbol{\theta}}^{\rm H}(\boldsymbol{R} + \boldsymbol{V}) \bar{\boldsymbol{\theta}} \label{optimal.7.a}\\
& ~~\textrm{s.t.} \quad~  |\theta_n| = 1, \forall n = 1,\ldots, N, \label{optimal.7.b}
\end{align}
\end{subequations}
where $ \bar{\boldsymbol{\theta}} = \begin{bmatrix} \boldsymbol{\theta} \\ t \end{bmatrix}$, $\boldsymbol{R} = \begin{bmatrix} \rho^2\boldsymbol{D}_h^{\rm H}\boldsymbol{G}^{\rm H} \boldsymbol{G} \boldsymbol{D}_h &
\rho \boldsymbol{D}_h^{\rm H}\boldsymbol{G}^{\rm H}\boldsymbol{h}_d \\
 \rho \boldsymbol{h}_d^{\rm H}\boldsymbol{G} \boldsymbol{D}_h & 0 \end{bmatrix}$,
and $\boldsymbol{V} = \begin{bmatrix} \rho(1-\rho)\textrm{diag}\{\boldsymbol{v}\} & 0 \\ 0 & 0 \end{bmatrix} $.
Note that
$\bar{\boldsymbol{\theta}}^{\rm H}(\boldsymbol{R} + \boldsymbol{V}) \bar{\boldsymbol{\theta}} = \textrm{tr}[(\boldsymbol{R} + \boldsymbol{V})\boldsymbol{Q}]$,
where $ \boldsymbol{Q} = \bar{\boldsymbol{\theta}}\bar{\boldsymbol{\theta}}^{\rm H} $. Clearly, $\boldsymbol{Q}$ is a positive semidefinite matrix, i.e., $\boldsymbol{Q}\succcurlyeq 0$, and $\text{rank}(\boldsymbol{Q})=1$. By relaxing the rank-one constraint on $\boldsymbol{Q}$, we can convert \eqref{optimal.7} into
\begin{align} \label{optimal.8}
& \max_{\boldsymbol{Q}} \quad \textrm{tr}( (\boldsymbol{R}+\boldsymbol{V})\boldsymbol{Q} ) \notag\\
& ~~\textrm{s.t.} \quad~  \boldsymbol{Q} \succcurlyeq 0; Q_{n,n}=1, \forall n = 1,\ldots, N+1.
\end{align}

The above problem is a standard SDP, and can be optimally solved by existing convex optimization solvers such as CVX \cite{grant2014cvx}.
The optimal $\boldsymbol{Q}$ of the SDP problem in \eqref{optimal.8} is not guaranteed to be rank-one in general. To obtain a suboptimal solution of $\bar{\boldsymbol{\theta}}$ from $\boldsymbol{Q}$, we follow \cite{so2007approximating} to take the eigenvalue decomposition of $\boldsymbol{Q}$ as
$\boldsymbol{Q} = \boldsymbol{U}\boldsymbol{\Sigma}\boldsymbol{U}^{\rm H}$, where $\boldsymbol{U} \in \mathbb{C}^{(N+1) \times (N+1)}$ is a unitary matrix and $\boldsymbol{\Sigma}\in \mathbb{C}^{(N+1) \times (N+1)}$ is a diagonal matrix.
Then, we obtain a suboptimal solution of $\bar{\boldsymbol{\theta}}$ as
$\bar{\boldsymbol{\theta}} = \boldsymbol{U}\boldsymbol{\Sigma}^{1/2}\boldsymbol{r}$, where $\boldsymbol{r} \in \mathbb{C}^{(N+1)} $ is a random vector
with each element generated from the circularly symmetric complex Gaussian (CSCG) distribution $\mathcal{CN}(0, 1)$. Then, the suboptimal solution of $\boldsymbol{\theta}$ in \eqref{optimal.2} is given by
$\boldsymbol{\theta} = \frac{\left[\bar{\boldsymbol{\theta}}\right]_{(1:N)}/\bar{\theta}_{N+1}}{\left\|\left[\bar{\boldsymbol{\theta}}\right]_{(1:N)}/\bar{\theta}_{N+1}\right\|_2}$, where $[\boldsymbol{a}]_{(1:N)}$ denotes the vector that contains the first $N$ elements of $\boldsymbol{a}$.

\section{Receiver Design}

\subsection{Problem Description}

The receiver aims to retrieve both the information from the user and sensors (i.e., $\boldsymbol{x}$ and $\boldsymbol{s}$).
More specifically, we rewrite \eqref{channel.H1} as
\begin{align} \label{channel.H3}
&\boldsymbol{Y}  = (\boldsymbol{A}\boldsymbol{s} + \boldsymbol{h}_d ) \boldsymbol{x}^{\rm T} + \boldsymbol{W}= \boldsymbol{z} \boldsymbol{x}^{\rm T} + \boldsymbol{W},
\end{align}
where $\boldsymbol{A}= \beta\boldsymbol{G} \boldsymbol{\Theta}\boldsymbol{D}_h \in \mathbb{C}^{M \times N}$ is a known coefficient matrix and
$\boldsymbol{z} = [z_1,z_2,\ldots,z_M]$ with $z_m = \boldsymbol{a}_m^{\rm H}\boldsymbol{s} + h_{d,m}$ and $\boldsymbol{a}_m^{\rm H}$ being the $m$-th row of $\boldsymbol{A}$. Given the algebraic structure between $\boldsymbol{Y}$ and $(\boldsymbol{x},\boldsymbol{s})$ in \eqref{channel.H2}, we propose the following two-step approach for the retrieval of $\boldsymbol{x}$ and $\boldsymbol{s}$: First recover $\boldsymbol{x}$ and $\boldsymbol{z}$ from
$\boldsymbol{Y}$, and then recover $\boldsymbol{s}$ from the recovered $\boldsymbol{z}$.\footnote{We emphasize that the two-step approach presented here is not necessarily optimal. However, we will show by numerical results that the performance loss due to the suboptimality of this two-step approach is usually marginal. }The details are presented in the following two subsections.

\subsection{Recovery of $\boldsymbol{x}$ and $\boldsymbol{z}$ }
 The recovery of $\boldsymbol{z}$ and $\boldsymbol{x}$ from the observation matrix $\boldsymbol{Y}$ can be regarded as a rank-$1$ matrix decomposition problem. We propose two methods, namely the singular value decomposition (SVD) method and the bilinear generalized approximate message passing (BiG-AMP)\cite{BiG-AMP} method, as detailed below.
\begin{itemize}
  \item {{\it SVD method:}}
      Let the SVD of $\boldsymbol{Y}$ be $\boldsymbol{Y}  = \boldsymbol{U}\boldsymbol{\Lambda}\boldsymbol{V}^{\rm H}$,
      where $\boldsymbol{U}= [\boldsymbol{u}_1, \boldsymbol{u}_2,\ldots,\boldsymbol{u}_M]$ and
      $\boldsymbol{V}= [\boldsymbol{v}_1, \boldsymbol{v}_2,\ldots,\boldsymbol{v}_M]$ are both unitary matrixes, and
      $\boldsymbol{\Lambda}= \diag\{\lambda_1, \lambda_2,\ldots,\lambda_M\}$ is a
      diagonal matrix with the elements in the diagonal sorted in a descending order, i.e.,
      $\lambda_1 \geqslant \lambda_2 \geqslant \ldots \geqslant \lambda_M$. We simply take the first column of $\boldsymbol{V}$ as an estimates of $\boldsymbol{x}$. That is, $\hat{\boldsymbol{x}}= \boldsymbol{v}_1$. The corresponding estimate of $\boldsymbol{z}$ is given by $\hat{\boldsymbol{z}}=\lambda_1\boldsymbol{u}_1$.
  \item {{\it BiG-AMP method:}} The BiG-AMP algorithm\cite{BiG-AMP} can be used to solve the factorization of $\boldsymbol{x}$ and $\boldsymbol{z}$
      from $\boldsymbol{Y}$. Note that the BiG-AMP algorithm requires the prior distributions of $\boldsymbol{z}$ and $\boldsymbol{x}$. We assume that the entries of $\boldsymbol{x}$ are independently and uniformly distributed over $\mathbb{C}$. As for $\boldsymbol{z}$, we approximate $z_m$, $\forall m$ as a CSCG random variable with the mean and the variance given by $\rho\boldsymbol{a}_m^{\rm H}\textbf{1} + h_{d,m}$ and $\rho(1-\rho)\|\boldsymbol{a}_m^{\rm H}\|_2^2$, respectively, where $\|\boldsymbol{a}\|_2$ denotes the $\ell_2-$norm of vector $\boldsymbol{a}$. Similarly to the SVD method, we denote the output estimates of $\boldsymbol{z}$ and $\boldsymbol{x}$ by $\hat{\boldsymbol{z}}$ and $\hat{\boldsymbol{x}}$, respectively.
\end{itemize}

There exists a scalar offset $\gamma$ in $\hat{\boldsymbol{z}}$ and $\hat{\boldsymbol{x}}$ since if $(\hat{\boldsymbol{z}}, \hat{\boldsymbol{x}})$ is a solution to \eqref{channel.H2}, then $(\hat{\boldsymbol{z}}/\gamma, \gamma\hat{\boldsymbol{x}})$ is also a valid solution to \eqref{channel.H2}\cite{Zhangjianwen}. The scalar offset can be eliminated by inserting a reference symbol in the first position of $\boldsymbol{x}$.
With the knowledge of the reference symbol $x_1$, $\gamma$ can be estimated by
$\gamma = x_1/\hat{x}_1$.
Then, the estimates of $\boldsymbol{z}$ and $\boldsymbol{x}$ are corrected as $\hat{\boldsymbol{z}}/\gamma$ and $\gamma \hat{\boldsymbol{x}}$, respectively. Finally, we map $\gamma \hat{\boldsymbol{x}}$ to the constellation of $\boldsymbol{x}$ as
\begin{align} \label{Rec.x}
\tilde{x}_i = \arg \min_{c \in \mathcal{C}} | c - \gamma \hat{x_i} |^2, \quad i = 1,\ldots, N.
\end{align}
\subsection{Recovery of $\boldsymbol{s}$ from $\boldsymbol{z}$ }

With $\tilde{\boldsymbol{x}}$ from \eqref{Rec.x}, we obtain an estimate of $\boldsymbol{z}$ as
\begin{align} \label{Rec.z}
\tilde{\boldsymbol{z}} = \frac{1}{LP}\boldsymbol{Y} \tilde{\boldsymbol{x}}^{*} = \boldsymbol{A}\boldsymbol{s} +  \boldsymbol{h}_d + \boldsymbol{w},
\end{align}
where $\boldsymbol{w} \in \mathbb{C}^{M \times 1}$ is a distortion term. Note that $\boldsymbol{h}_d$ can be precancelled from $\tilde{\boldsymbol{z}}$
prior to the recovery of $\boldsymbol{s}$.
Also note that $\boldsymbol{w}$ is an additive white Gaussian noise (AWGN) with the elements independent and identically distributed drawn from $\mathcal{CN}(0,\sigma_w^2/P)$ when $\tilde{\boldsymbol{x}} = \boldsymbol{x}$. In general, $\tilde{\boldsymbol{x}}$ may contain errors, and so the actual power of $\boldsymbol{w}$ is slightly higher than $\sigma_w^2/P$.

The recovery of $\boldsymbol{s}$ from $\tilde{\boldsymbol{z}}$ can be done by noting that $\boldsymbol{s}$ is a highly structured signal composed of only $0$s and $1$s. Structured signal recovery algorithms, such as orthogonal matching pursuit (OMP)\cite{tropp2007signal} and compressive sampling matching pursuit (CoSaMP)\cite{needell2009cosamp}, can be used to recovery of $\boldsymbol{s}$ from $\boldsymbol{z}$ in \eqref{Rec.z}.
In order to make the best use of the prior knowledge of $\boldsymbol{s}$, we use the generalized approximate message passing (GAMP) algorithm \cite{rangan2011generalized} for the recovery of $\boldsymbol{s}$. The details of the algorithm is omitted due to space limitation.

\section{Numerical Results}

In simulations, the entries of $\boldsymbol{x}$ are QPSK modulated with Gray-mapping.
Following \cite{nadeem2019large, wu2018intelligent, huang2018large},
we generate the entries of $\boldsymbol{G}$, $\boldsymbol{h}_r$, and $\boldsymbol{h}_d $ independently from the CSCG distribution $\mathcal{CN}(0, 1)$.
 We set the amplitude reflection coefficient $\beta=0.5$ and the transmission power at the user $P = 1$. The SNR is defined as $\text{SNR}=1/\sigma_w^2$.
The maximum numbers of BiG-AMP and GAMP iterations are set to $200$ and $50$, respectively.
The simulation results presented in this paper are obtained by taking average over 5000 random realizations.
For the recovery of $\boldsymbol{x}$, we compare the approaches listed  below.
\begin{itemize}
  \item { {Without LIS:}} Recover $\boldsymbol{x}$ from $\boldsymbol{Y}$ without the enhancement of LIS, i.e., $\boldsymbol{s}=\textbf{0}$
  \item {{SVD:}} The SVD based method proposed in this paper.
  \item {{BiG-AMP:}} The BiG-AMP based method proposed in this paper.
  \item { {LB-$\boldsymbol{x}$:}} The lower bound of $\boldsymbol{x}$, in which $\boldsymbol{x}$ is estimated under the perfect knowledge of $\boldsymbol{s}$ as
  $\tilde{x}_i = \arg \min_{c \in \mathcal{C}} | c - (\boldsymbol{z}^{\rm H} \boldsymbol{Y})_i|^2 $, for $i=1,\ldots,N$.
\end{itemize}
For the recovery of $\boldsymbol{s}$, we compare the approaches listed  below.
\begin{itemize}
  \item { {SVD+GAMP:}} First recover $\boldsymbol{z}$ by using the SVD method, and then recover $\boldsymbol{s}$ from the estimated $\boldsymbol{z}$ using the GAMP algorithm.
  \item { {BiG-AMP+GAMP:}} First recover $\boldsymbol{z}$ by using the BiG-AMP method, and then recover $\boldsymbol{s}$ from the estimated $\boldsymbol{z}$ using the GAMP algorithm.
  \item { {BiG-AMP+OMP:}} First recover $\boldsymbol{z}$ by using the BiG-AMP method, and then recover $\boldsymbol{s}$ from the estimated $\boldsymbol{z}$ using the OMP algorithm.
  \item { {BiG-AMP+CoSaMP:}} First recover $\boldsymbol{z}$ by using the BiG-AMP method, and then recover $\boldsymbol{s}$ from the estimated $\boldsymbol{z}$ using the CoSaMP algorithm.
  \item { {LB-$\boldsymbol{s}$:}} The lower bound of $\boldsymbol{s}$, in which $\boldsymbol{s}$ is estimated by the GAMP algorithm with perfectly known $\boldsymbol{x}$.
\end{itemize}

\begin{figure}
  \centering
  \includegraphics[width=3.5 in]{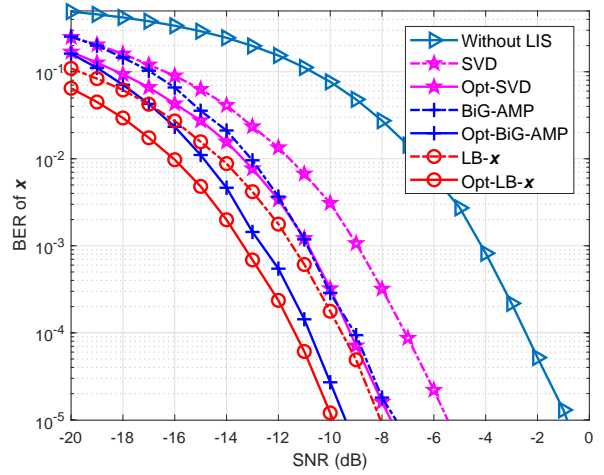}\\
  \caption{The average BER of $\boldsymbol{x}$ versus SNR for different approaches. $M=32$, $N=32$, $L=100$, and $\rho=0.5$.}\label{X_BER}
\end{figure}
\begin{figure}
  \centering
  \includegraphics[width=3.5 in]{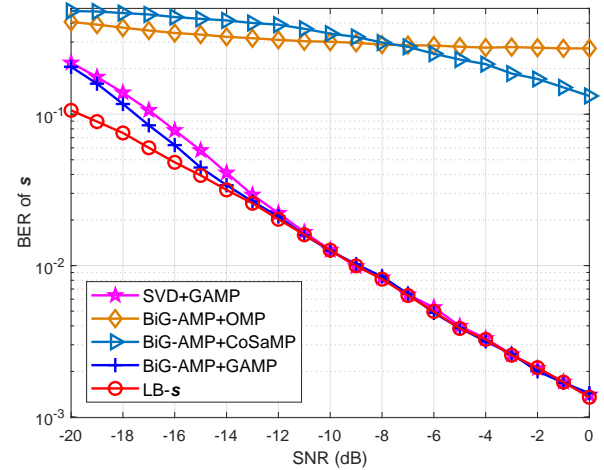}\\
  \caption{The average BER of $\boldsymbol{s}$ versus SNR for different approaches. $M=32$, $N=32$, $L=100$, and $\rho=0.5$.}\label{S_BER}
\end{figure}

Fig.~\ref{X_BER} compares the average bit error rate (BER) of $\boldsymbol{x}$ versus SNR for the GAMP algorithm without LIS, the SVD approach, the BiG-AMP approach, and the lower bound LB-$\boldsymbol{x}$.  The dotted lines are generated with random phase shift $\boldsymbol{\Theta}$; the solid lines are generated with optimized phase shift $\boldsymbol{\Theta}$. The prefix ``Opt-'' means that the corresponding curve is obtained by using optimized phase shifts.
The other settings are $M=32$, $N=32$, $L=100$, and $\rho=0.5$. Fig.~\ref{X_BER} shows that by optimization $\boldsymbol{\Theta}$, the system achieves about $2$ dB SNR improvement when the antenna activity rate of the LIS is $\rho = 0.5$.
From Fig.~\ref{X_BER}, we see that with the enhance of the LIS, the system performance can be improved by about $5$ dB to $9$ dB at $\text{BER}=10^{-5}$ for the various schemes under consideration. We also see that the BiG-AMP method approaches the lower bound to within $0.5$ dB at $\text{BER}=10^{-5}$
and the SVD method has an SNR gap of about $2$ dB SNR gap away from the BiG-AMP method.
\begin{figure}
  \centering
  \includegraphics[width=3.5 in]{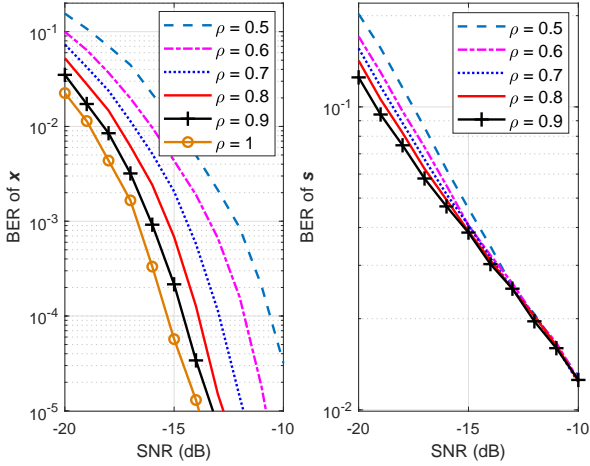}\\
  \caption{The average BER of $\boldsymbol{x}$ and $\boldsymbol{s}$ versus SNR with $\rho\in \{0.5, 0.6, 0.7, 0.8,0.9,1$\}. $M=32$, $N=32$, $L=100$.}\label{Sparsity}
\end{figure}

Fig.~\ref{S_BER} compares the average BER of $\boldsymbol{s}$ versus SNR for the GAMP approach, the OMP approach, the CoSaMP approach, and the lower bound LB-$\boldsymbol{s}$. We see that the GAMP approach achieves the lower bound at the SNR as low as $-14$ dB. We also see that the BER of $\boldsymbol{s}$ for BiG-AMP+GAMP slightly outperforms that of the SVD+GAMP. This is because the BiG-AMP algorithm performs better than the SVD algorithm in the recovery of $\boldsymbol{z}$. By exploiting the prior information of $\boldsymbol{s}$, the GAMP algorithm significantly outperforms the OMP algorithm and the CoSaMP algorithm.

Fig.~\ref{Sparsity} compares the average BER of $\boldsymbol{x}$ and $\boldsymbol{s}$ versus SNR with $\rho$ varying from $0.5$ to $1$. The other settings are $M=32$, $N=32$, and $L=100$.
 We see that with the increase of $\rho$, both the BERs of $\boldsymbol{x}$ and $\boldsymbol{s}$ decrease due to the increase of the receive SNR. However, the cost is the information rate of $\boldsymbol{s}$. Specifically, the information rate per entry of $\boldsymbol{s}$ is $1.0000$, $0.9710$, $0.8813$, $0.7219$, $0.4690$, and $0$ for $\rho=0.5$, $0.6$, $0.7$, $0.8$, $0.9$, and $1$, respectively.
We see that the rate of $\boldsymbol{s}$ increase from $0$ to $0.4690$ by reducing $\rho$ from $1$ to $0.9$, at the cost of only about $0.5$ dB of SNR loss at BER of $\boldsymbol{x}=10^{-5}$.
This demonstrates an attractive tradeoff between the recovery performance and the information rate of the LIS.

\section{Conclusions}

In this paper, we proposed a PBIT enhanced wireless system, in which a LIS simultaneously enhances the
user-BS communication (by adjusting the phases of reflected electro-magnetic waves on the activated reflecting elements of the LIS) and transmits information to the receiver (by modulating the on-off states of the reflecting elements of the LIS).
We further proposed to optimize the phase shift vector $\boldsymbol{\theta}$ to maximize the average receiver SNR. An efficient algorithm was developed to approximately solve the problem. We also proposed a two-step approach to the retrieval of the LIS signal $\boldsymbol{s}$ and the user signal $\boldsymbol{x}$. Substantial performance gains have been demonstrated for our proposed design.



\begin{thebibliography}{50}
\providecommand{\url}[1]{#1}
\csname url@samestyle\endcsname
\providecommand{\newblock}{\relax}
\providecommand{\bibinfo}[2]{#2}
\providecommand{\BIBentrySTDinterwordspacing}{\spaceskip=0pt\relax}
\providecommand{\BIBentryALTinterwordstretchfactor}{4}
\providecommand{\BIBentryALTinterwordspacing}{\spaceskip=\fontdimen2\font plus
\BIBentryALTinterwordstretchfactor\fontdimen3\font minus
  \fontdimen4\font\relax}
\providecommand{\BIBforeignlanguage}[2]{{%
\expandafter\ifx\csname l@#1\endcsname\relax
\typeout{** WARNING: IEEEtran.bst: No hyphenation pattern has been}%
\typeout{** loaded for the language `#1'. Using the pattern for}%
\typeout{** the default language instead.}%
\else
\language=\csname l@#1\endcsname
\fi
#2}}
\providecommand{\BIBdecl}{\relax}
\BIBdecl

\bibitem{boccardi2013five}
F.~Boccardi, R.~W. Heath~Jr, A.~Lozano, T.~L. Marzetta, and P.~Popovski, ``Five
  disruptive technology directions for 5{G},'' \emph{arXiv preprint
  arXiv:1312.0229}, 2013.

\bibitem{andrews2014will}
J.~G. Andrews, S.~Buzzi, W.~Choi, S.~V. Hanly, A.~Lozano, A.~C. Soong, and
  J.~C. Zhang, ``What will {5G} be?'' \emph{IEEE J. Sel. Areas Commun.},
  vol.~32, no.~6, pp. 1065--1082, Jun. 2014.

\bibitem{zhang2017fundamental}
S.~Zhang, Q.~Wu, S.~Xu, and G.~Y. Li, ``Fundamental green tradeoffs:
  Progresses, challenges, and impacts on {5G} networks,'' \emph{IEEE Commun.
  Surveys Tuts.}, vol.~19, no.~1, pp. 33--56, Jul. 2017.

\bibitem{nadeem2019large}
Q.-U.-A. Nadeem, A.~Kammoun, A.~Chaaban, M.~Debbah, and M.-S. Alouini, ``Large
  intelligent surface assisted {MIMO} communications,'' \emph{arXiv preprint
  arXiv:1903.08127}, 2019.

\bibitem{wu2018intelligent}
Q.~Wu and R.~Zhang, ``Intelligent reflecting surface enhanced wireless network
  via joint active and passive beamforming,'' \emph{arXiv preprint
  arXiv:1810.03961}, 2018.

\bibitem{hum2014reconfigurable}
S.~V. Hum and J.~Perruisseau-Carrier, ``Reconfigurable reflectarrays and array
  lenses for dynamic antenna beam control: A review,'' \emph{IEEE Trans.
  Antennas Propag.}, vol.~62, no.~1, pp. 183--198, Oct. 2013.

\bibitem{foo2017liquid}
S.~Foo, ``Liquid-crystal reconfigurable metasurface reflectors,'' in
  \emph{proc. 2017 IEEE ISAP}, 2017, pp. 2069--2070.

\bibitem{huang2018large}
C.~Huang, A.~Zappone, G.~C. Alexandropoulos, M.~Debbah, and C.~Yuen, ``Large
  intelligent surfaces for energy efficiency in wireless communication,''
  \emph{arXiv preprint arXiv:1810.06934}, 2018.

\bibitem{mesleh2008spatial}
R.~Y. Mesleh, H.~Haas, S.~Sinanovic, C.~W. Ahn, and S.~Yun, ``Spatial
  modulation,'' \emph{IEEE Trans. Veh. Technol.}, vol.~57, no.~4, pp.
  2228--2241, Jul. 2008.

\bibitem{zhang2019constellation}
Q.~Zhang, H.~Guo, Y.-C. Liang, and X.~Yuan, ``Constellation learning-based
  signal detection for ambient backscatter communication systems,'' \emph{IEEE
  J. Sel. Areas Commun.}, vol.~37, no.~2, pp. 452--463, Sep. 2019.

\bibitem{green1963general}
R.~B. Green, ``{T}he {G}eneral {T}heory of {A}ntenna {S}cattering,'' Ph.D.
  dissertation, Ohio State Univ., Columbus, 1963.

\bibitem{he2019cascaded}
Z.-Q. He and X.~Yuan, ``Cascaded channel estimation for large intelligent
  metasurface assisted massive {MIMO},'' \emph{arXiv preprint
  arXiv:1905.07948}, 2019.

\bibitem{taha2019enabling}
A.~Taha, M.~Alrabeiah, and A.~Alkhateeb, ``Enabling large intelligent surfaces
  with compressive sensing and deep learning,'' \emph{arXiv preprint
  arXiv:1904.10136}, 2019.

\bibitem{cover2012elements}
T.~M. Cover and J.~A. Thomas, \emph{Elements of information theory}.\hskip 1em
  plus 0.5em minus 0.4em\relax John Wiley \& Sons, 2012.

\bibitem{zhang2017multi}
J.~Zhang, T.~Wei, X.~Yuan, and R.~Zhang, ``Multi-antenna constant envelope
  wireless power transfer,'' \emph{IEEE Trans. Green Commu. and Networking},
  vol.~1, no.~4, pp. 458--467, Aug. 2017.

\bibitem{grant2014cvx}
M.~Grant and S.~Boyd, ``{CVX}: Matlab software for disciplined convex
  programming, version 2.1,'' 2014.

\bibitem{so2007approximating}
A.~M.-C. So, J.~Zhang, and Y.~Ye, ``On approximating complex quadratic
  optimization problems via semidefinite programming relaxations,''
  \emph{Mathematical Programming}, vol. 110, no.~1, pp. 93--110, Jun. 2007.

\bibitem{BiG-AMP}
J.~T. Parker, P.~Schniter, and V.~Cevher, ``Bilinear generalized approximate
  message passing part {I}: Derivation,'' \emph{IEEE Trans. Signal Process.},
  vol.~62, no.~22, pp. 5839--5853, Nov. 2014.

\bibitem{Zhangjianwen}
J.~Zhang, X.~Yuan, and Y.~Zhang, ``Blind signal detection in massive {MIMO}:
  Exploiting the channel sparsity,'' \emph{IEEE Trans. Commun.}, vol.~66,
  no.~2, pp. 700--712, Feb. 2018.

\bibitem{tropp2007signal}
J.~A. Tropp and A.~C. Gilbert, ``Signal recovery from random measurements via
  orthogonal matching pursuit,'' \emph{IEEE Trans. Inf. Theory}, vol.~53,
  no.~12, pp. 4655--4666, 2007.

\bibitem{needell2009cosamp}
D.~Needell and J.~A. Tropp, ``Cosamp: Iterative signal recovery from incomplete
  and inaccurate samples,'' \emph{Appl. Comput. Harmonic Anal.}, vol.~26,
  no.~3, pp. 301--321, May 2009.

\bibitem{rangan2011generalized}
S.~Rangan, ``Generalized approximate message passing for estimation with random
  linear mixing,'' in \emph{proc. IEEE Int. Symp. Inf. Theory}, 2011, pp.
  2168--2172.

\end{thebibliography}

\end{document}